\documentclass[12pt,preprintnumbers,nofootinbib,letterpaper]{article}
\pdfoutput=1
\usepackage{amsmath,amssymb,amsfonts,cite,xcolor,epsf,epsfig,epstopdf,graphics,booktabs,graphicx,mathtools,subcaption,physics,verbatim}
\def\thefootnote{*\arabic{footnote}}
\definecolor{ultramarine}{rgb}{0.07, 0.04, 0.56}
\definecolor{cadmiumgreen}{rgb}{0.0, 0.42, 0.24}
\definecolor{indigo(dye)}{rgb}{0.0, 0.25, 0.42}
\usepackage[linktocpage=true,breaklinks]{hyperref}
\hypersetup{
	colorlinks=true,
	citecolor=ultramarine,
	linkcolor=cadmiumgreen,
	urlcolor=indigo(dye),
}

\usepackage[utf8]{inputenc}

\numberwithin{equation}{section}

\usepackage{array}
\newcolumntype{P}[1]{>{\centering\arraybackslash}p{#1}}

\usepackage[shortlabels]{enumitem}

\usepackage{dcolumn}
\newcolumntype{M}[1]{>{\centering\arraybackslash}m{#1}}
\newcolumntype{N}{@{}m{0pt}@{}}

\usepackage{amsmath}
\usepackage{ulem}

\usepackage[utf8]{inputenc}
\usepackage[T1]{fontenc}
\usepackage{lmodern}
\usepackage{geometry}
\usepackage{amsmath,amssymb,amsthm,mathtools}
\usepackage{bm}
\usepackage{microtype}
\usepackage{physics}

\newcommand{\D}{{\rm d}}

\newcommand{\be}{\begin{equation}}  
	\newcommand{\ee}{\end{equation}}

\usepackage{tikz}

\begin{document}
	

	\begin{center}
		
		\def\thefootnote{\fnsymbol{footnote}}
		
		\vspace*{1.5cm}
		{\Large {\bf Fragility of stealth solutions in mimetic gravity}}
		\\[1cm]
		
		{Alireza Allahyari$^{1}$,
			Clarisse Donio$^{2}$,
			Mohammad Ali Gorji$^{3}$,
			\\
			Reera Gorji$^{4,5}$, Javad T. Firouzjaee$^{4,5}$}
		\\[.7cm]
		
		{\small\textit{$^{1}$Department of Astronomy and High Energy Physics, Kharazmi University, 15719-14911, Tehran, Iran}}\\
		\vspace{0.25cm}
		{\small\textit{$^{2}$Magist\`ere de Physique Fondamentale d’Orsay, Universit\'e Paris-Saclay, 91400 Orsay, France}}\\
		\vspace{0.25cm}
		{\small\textit{$^{3}$Cosmology, Gravity, and Astroparticle Physics Group, Center for Theoretical Physics of the Universe, Institute for Basic Science (IBS), Daejeon, 34126, Korea}}\\
		\vspace{0.25cm}
		{\small\textit{$^{4}$Department of Physics, K.N. Toosi University of Technology, P.O. Box 15875-4416, Tehran, Iran}}\\
		\vspace{0.25cm}
		{\small\textit{$^{5}$PDAT Laboratory, Department of Physics, K. N. Toosi University of Technology, P.O. Box 15875-4416, Tehran, Iran}}
		
	\end{center}
	
	\vspace{1cm}
	
	\hrule \vspace{0.5cm}
	
	\begin{abstract}
		We study a broad class of constrained mimetic-type extensions of general relativity with action $S=\int\D^4x\sqrt{-g}\,\bigl(R/2+\lambda\,C[g,\Psi]+{\cal L}_{\rm m}\bigr)$, where $R$ is the Ricci scalar, $\lambda$ is a Lagrange multiplier, $C[g,\Psi]$ is a scalar functional of the metric and generic field content $\Psi$ (possibly involving $\Psi$ and its covariant derivatives) and ${\cal L}_{\rm m}$ is the matter Lagrangian. On the branch $\bar{\lambda}=0$, with the bar denoting a background value, the constrained sector drops out of the background dynamics, so, on domains where a background profile $\bar\Psi$ satisfying $\bar C=0$ exists, the theory admits the corresponding general relativity geometries as stealth solutions. As an explicit illustration, we consider the scalar field case, where $C=g^{\mu\nu}\partial_\mu\phi\partial_\nu\phi\pm1=0$ becomes a Hamilton-Jacobi equation selecting geodesic congruences. In this setting, we discuss spherically symmetric and rotating stealth profiles using Carter separability. We then analyze perturbations at the general level. The quadratic terms weighted by $\bar\lambda$ vanish on the stealth branch, while the nonlinear constraint still holds order by order and nonlinear interactions need not vanish. Consequently, reproducing GR at the background level does not ensure a uniform $\bar\lambda\to0$ limit for perturbations, and strong coupling may arise.
	\end{abstract}
	\hrule
	\def\thefootnote{\arabic{footnote}}
	\setcounter{footnote}{0}
	

	
	\newpage
	
	\section{Introduction}\label{sec-Intro}
	
	General relativity (GR) provides an exceptionally successful description of gravitational phenomena across a wide range of scales, from laboratory and solar system tests to strong field observations around compact objects. At the same time, a number of open problems in gravitational physics and cosmology motivate exploring extensions of GR, including scenarios in which additional degrees of freedom are present but their effects can be suppressed in regimes where GR is well tested \cite{Jain:2010ka,Brax:2013ida,Kobayashi:2014ida,Joyce:2014kja,Koyama:2015vza,Joyce:2016vqv,Burrage:2017qrf,Crisostomi:2017lbg,Dima:2017pwp,Ishak:2018his,Amendola:2019laa,Kobayashi:2019hrl,Ferreira:2019xrr,Sakstein:2018fwz,Brax:2021wcv,Fischer:2024eic}. A common theme in such modified gravity frameworks is therefore scale dependence: the extra sector may play an important role on large scales, while local and astrophysical environments must remain close to GR. In many theories this is achieved through a screening mechanism (for instance Vainshtein-type screening in certain scalar-tensor models \cite{Kimura:2011dc,Koyama:2015vza,Kobayashi:2019hrl}), ensuring that deviations from GR become small in the vicinity of astrophysical sources.
	
	Mimetic-type constructions provide a distinct class of constrained extensions of GR. Originally introduced as a singular disformal transformation of the metric \cite{Chamseddine:2013kea,Deruelle:2014zza,Arroja:2015wpa}, mimetic models can be formulated more generally as constrained extensions of GR in which a Lagrange multiplier enforces a condition on the kinetic structure of additional field(s) \cite{Golovnev:2013jxa,Barvinsky:2013mea,Chamseddine:2014vna,Chaichian:2014qba,Mirzagholi:2014ifa,BenAchour:2016cay,Sebastiani:2016ras,Firouzjahi:2017txv,Zheng:2017qfs,Takahashi:2017pje,Hirano:2017zox,Gorji:2017cai,Vagnozzi:2017ilo,Langlois:2018jdg,Firouzjahi:2018xob,Gorji:2018okn,Jirousek:2018ago,Casalino:2018tcd,Gorji:2019ttx,Ganz:2019vre,Gorji:2020ten,Mansoori:2021fjd,Jirousek:2022rym,Zheng:2022vwm,Jirousek:2022kli,Domenech:2023ryc,Visa:2024fii,Colleaux:2025vtm,Gorji:2025ajb,Gorji:2025fbv}. We consider a general mimetic-type theory\footnote{We work with signature $-+++$ and set $8\pi G=1$.}
	\begin{equation}\label{eq:action}
		S \;=\; \int \D^4x\,\sqrt{-g}\left[\frac{1}{2}R+\lambda\,C[g,\Psi]+{\cal L}_{\rm m}[g,\Xi]\right],
	\end{equation}
	where $R$ is the Ricci scalar, $C[g,\Psi]$ is a scalar functional built from the metric $g_{\mu\nu}$ and a set of fields $\Psi$ (which may include scalars, vectors, gauge fields, or more general tensor fields), possibly depending on $\Psi$ and a finite number of its covariant derivatives and ${\cal L}_{\rm m}$ is the Lagrangian of the matter field $\Xi$. This formulation includes the usual scalar mimetic model as the special case $\Psi=\phi$ with
	$C=g^{\mu\nu}\partial_\mu\phi\partial_\nu\phi\pm1$ \cite{Chamseddine:2013kea,Golovnev:2013jxa,Gorji:2020ten}. It also includes vector realizations, with $\Psi=X_\mu$ and $C=g^{\mu\nu}X_\mu X_\nu\pm1$ \cite{Barvinsky:2013mea,Chaichian:2014qba}, as well as gauge-field realizations, such as $C=X_{\mu\nu}X^{\mu\nu}\pm1$ with $X_{\mu\nu}=\partial_\mu X_\nu-\partial_\nu X_\mu$ \cite{Gorji:2018okn}. More generally, one may consider other
	choices of field content and constraint functionals $C[g,\Psi]$, including vector and higher-rank tensor
	extensions, as well as constraints involving covariant derivatives of $\Psi$ \cite{Gorji:2018okn,Jirousek:2018ago,Gorji:2019ttx}.\footnote{A similar Lagrange-multiplier structure appears in Einstein-\ae ther theory \cite{Jacobson:2000xp} and in the covariant formulation of the infrared limit of Ho\v{r}ava-Lifshitz gravity \cite{Blas:2009yd}. Our setup \eqref{eq:action} thus overlaps with the constrained sector of such models, though their standard formulations include additional derivative interactions that can modify the background branches and perturbation dynamics. Extending our analysis to these theories is interesting but beyond the scope of this work.}

	In cosmological settings, these models are often invoked to address large-scale phenomena ranging from dark matter-like behavior \cite{Chamseddine:2013kea} to late-time acceleration \cite{Jirousek:2018ago,Hammer:2020dqp} and other effective modifications \cite{Gorji:2018okn,Gorji:2019ttx}, where the Lagrange multiplier $\lambda$ typically acquires a non-vanishing background value. In the scalar realization, the multiplier is proportional to the energy density associated with the constrained sector. Motivated by these applications, here we study local GR backgrounds with $\bar\lambda=0$ and deviations from them for nonzero $\bar\lambda$. In the scalar case, the curvature scale sourced by the constrained sector is of order $|\bar\lambda|$. For a compact object of mass $M$, a local deviation scale can be estimated by comparing this with the curvature scale of the Schwarzschild geometry, ${\cal K}^{1/2}\sim M/r^3$, where ${\cal K}=R_{\mu\nu\alpha\beta}R^{\mu\nu\alpha\beta}$ is the Kretschmann scalar. This gives
	\begin{align}\label{r-screen}
		r_{\mathrm{dev}}\sim\left(\frac{M}{|\bar\lambda(r_{\mathrm{dev}})|}\right)^{\frac13} \,.
	\end{align}
	In Appendix~\ref{app:screenradius}, we obtain the same scaling from a local, time-dependent spherical deformation of Schwarzschild in the scalar realization.
	
	We therefore focus on the exactly stealth branch, $\bar\lambda=0$. On this branch, the constrained sector drops out of the background dynamics, and any GR geometry is admitted as a solution provided the remaining fields $\bar\Psi$ can be chosen to satisfy the background constraint $\bar C=0$. These are stealth configurations: the fields $\bar\Psi$ may have nontrivial profiles, but they do not backreact on the metric. Thus Schwarzschild, Kerr, and all general GR backgrounds can arise as local solutions of the theory whenever the background constraint can be satisfied.
	
	The central result of this paper is that this apparently benign local branch is perturbatively subtle. Since the Lagrange multiplier enters the action linearly, the constrained sector enforces $C[g,\Psi]=0$ order by order in fluctuations. The quadratic terms weighted by $\bar\lambda$ disappear on the stealth branch, while nonlinear interactions can remain. If these $\bar\lambda$-weighted quadratic terms contribute to the kinetic operator of a physical mode after the constraints are resolved, their disappearance at $\bar\lambda=0$ can make the $\bar\lambda\to0$ limit non-uniform and may lead to strong coupling. Thus reproducing GR at the level of background solutions does not by itself establish a regular perturbative limit.

	In Sec.~\ref{sec-1}, we present the general framework and show that GR geometries arise as stealth solutions on the $\bar\lambda=0$ branch whenever the background constraint can be satisfied. In Sec.~\ref{sec-scalar}, we specialize to the scalar realization and summarize explicit stealth profiles on rotating and non-rotating spacetimes. In Sec.~\ref{sec-pert}, we analyze perturbations around stealth backgrounds in the general theory and examine the disappearance of the $\bar\lambda$-weighted quadratic terms and its implications for the $\bar\lambda\to0$ limit. Finally, Sec.~\ref{sec:summary} summarizes our results and discusses their implications for the viability of mimetic-type modified gravities admitting GR-like local regimes. Some technical details are presented in the appendices.

	\section{Stealth GR solutions}\label{sec-1}
	
	We consider the general constrained mimetic-type theory introduced in Sec.~\ref{sec-Intro}, with action \eqref{eq:action}.
	
	Variation with respect to $\lambda$ gives the constraint equation
    \begin{equation}\label{eq:constraint}
		C[g,\Psi]=0\,,
	\end{equation}
	where $C[g,\Psi]$ is a scalar functional of the metric $g_{\mu\nu}$ and the fields $\Psi$, possibly involving a finite number of their covariant derivatives. We keep this functional completely general.
	
	Variation of the action~\eqref{eq:action} with respect to the metric yields	\begin{equation}\label{eq:metric-eom-general-matter}
		\begin{gathered}
			G_{\mu\nu} = T^{\rm c}_{\mu\nu}+T^{\rm m}_{\mu\nu} 
			\,;
			\\
			T^{\rm c}_{\mu\nu} \equiv \frac{-2}{\sqrt{-g}}\frac{\delta}{\delta g^{\mu\nu}}	\!\left(\sqrt{-g}\,\lambda\,C[g,\Psi]\right) \,, \qquad
			T^{\rm m}_{\mu\nu} \equiv \frac{-2}{\sqrt{-g}}\frac{\delta}{\delta g^{\mu\nu}}	\!\left(\sqrt{-g}\,{\cal L}_{\rm m}[g,\Xi]\right) \,,
		\end{gathered}
	\end{equation}
	where $G_{\mu\nu} = R_{\mu\nu}-\frac{1}{2}Rg_{\mu\nu}$ is the Einstein tensor, $T^{\rm c}_{\mu\nu}$ denotes the effective stress tensor sourced by the constrained sector, with ${\rm c}$ standing for ``constraint", and $T^{\rm m}_{\mu\nu}$ is the energy-momentum tensor of the matter.
	
	Variation with respect to $\Psi$ gives the Euler-Lagrange equations
	\begin{equation}\label{eq:psi-eom-general}
		\mathcal{E}_{\Psi}\!\left(\lambda\,C[g,\Psi]\right)=0\,,
	\end{equation}
	where $\mathcal{E}_{\Psi}$ denotes the Euler-Lagrange operator appropriate to the field content. For example, for a vector field $\Psi=X_{\mu}$ with $C=X_{\mu\nu}X^{\mu\nu}\pm1$ and $X_{\mu\nu}=\partial_\mu X_\nu-\partial_\nu X_\mu$, this equation gives $\nabla_\mu(\lambda X^{\mu\nu})=0$ \cite{Gorji:2018okn,Gorji:2019ttx,Gorji:2025fbv}.
	
	Similarly, variation with respect to the matter field $\Xi$ gives
	\begin{equation}\label{eq:xi-eom-general}
		\mathcal{E}_{\Xi}\!\left({\cal L}_{\rm m}[g,\Xi]\right)=0\,.
	\end{equation}
	
	We are interested in stealth solutions on the branch $\bar\lambda=0$. In the following two subsections, we treat separately the cases with and without matter.
	
	\subsection{Solutions with matter}
	
	Let us first consider a background configuration of the form
	\begin{align}\label{eq:BG-config-matter}
		\{
		\bar{g}_{\mu\nu}\,, \bar{\Psi}\,, \bar{\lambda}=0\,,\bar{\Xi}
		\}\,,
	\end{align}
	where a bar denotes a background quantity. On this branch, the $\Psi$-equations \eqref{eq:psi-eom-general} are satisfied identically. Indeed, the constrained sector enters the action only through the product $\lambda\,C[g,\Psi]$, so the corresponding Euler-Lagrange equations \eqref{eq:psi-eom-general} contain terms proportional to $\lambda$ and its derivatives. Hence, for $\bar\lambda=0$, their background value vanishes independently of the explicit form of $C[g,\Psi]$.
	
	The constraint equation \eqref{eq:constraint} gives
	\begin{equation}\label{eq:constraint-BG}
		\bar C \equiv C[\bar g,\bar\Psi]=0\,.
	\end{equation}
	The only remaining nontrivial equations are therefore the metric equations. Since the constrained-sector contribution vanishes on the branch $\bar\lambda=0$, they reduce to
	\begin{align}\label{eq:EE-matter}
		{\bar G}_{\mu\nu}=\bar T^{\rm m}_{\mu\nu}\,.
	\end{align}
	Thus any GR solution with matter is also a solution of the constrained theory, provided one can choose a profile $\bar\Psi$ satisfying \eqref{eq:constraint-BG}. For example, choosing the minimally coupled matter sector to be electromagnetism, ${\cal L}_{\rm m}[g,\Xi]=-\frac{1}{4}F_{\mu\nu}F^{\mu\nu}$ with $F_{\mu\nu}=\partial_\mu A_\nu-\partial_\nu A_\mu$, gives the usual Reissner-Nordstr\"om geometry together with a possible nontrivial stealth profile $\bar\Psi$.
	
	\subsection{Vacuum solutions}
	
	The vacuum case follows in the same way, but with no matter fields. We consider
	\begin{align}\label{eq:BG-config}
		\{
		\bar{g}_{\mu\nu}\,, \bar{\Psi}\,, \bar{\lambda}=0
		\}\,.
	\end{align}
	As before, the $\Psi$-equations are automatically satisfied on this branch, while the auxiliary-field equation imposes the same background constraint \eqref{eq:constraint-BG}. The metric equations now reduce to the vacuum Einstein equations,
	\begin{align}\label{eq:EE-vacuum}
		{\bar R}_{\mu\nu}=0\,,
	\end{align}
	where $\bar{R}_{\mu\nu}$ is the background Ricci tensor constructed from ${\bar g}_{\mu\nu}$.
	
	Therefore every vacuum GR geometry, including Schwarzschild, Kerr, and more general Ricci-flat spacetimes, is admitted as a solution of the theory on the branch $\bar\lambda=0$, again provided the background fields $\bar\Psi$ can be chosen to satisfy $\bar C=0$. These are stealth configurations: the fields $\bar\Psi$ may have nontrivial profiles, but they do not backreact on the geometry.

In the following section we make the above general construction explicit in the scalar realization, where the constraint becomes a Hamilton-Jacobi equation and the connection with geodesic congruences makes the stealth mechanism particularly transparent. Other choices of the constrained field lead to different types of constraint equations. In particular, gauge-field realizations with constraints such as $C=X_{\mu\nu}X^{\mu\nu}\pm1$ do not lead to a Hamilton-Jacobi equation, since the constraint fixes an invariant of the field strength rather than the norm of a scalar gradient. Stealth black holes with Abelian and non-Abelian mimetic gauge fields were constructed in \cite{Gorji:2025fbv}. For this reason, here we focus on the scalar realization and study explicit examples on black hole spacetimes.

We now turn to these scalar examples, where the interplay between the background constraint \eqref{eq:constraint-BG} and spacetime symmetries can be studied in detail.

	\section{Example: scalar field}\label{sec-scalar}
	
	We consider the scalar-field realization of the constraint. This case is motivated by mimetic gravity and scalar-field dark matter models, where a scalar degree of freedom is constrained to mimic the behavior of pressureless dust or other matter components \cite{Chamseddine:2013kea}. Concretely, we take \cite{Golovnev:2013jxa}
	\begin{equation}\label{eq:C_scalar_choice}
		\Psi=\phi,
		\qquad
		C[g,\phi]=g^{\mu\nu}\partial_\mu\phi\,\partial_\nu\phi+\eta,
		\qquad
		\eta=\pm1.
	\end{equation}
	Substituting this in the general action~\eqref{eq:action}, we obtain the specific action given by
	\begin{equation}\label{eq:action_scalar}
		S=\int \D^4x\,\sqrt{-g}\left[
		\frac{1}{2}R+\lambda\left(g^{\mu\nu}\partial_\mu\phi\,\partial_\nu\phi+\eta\right)+{\cal L}_{\rm m}[g,\Xi]
		\right].
	\end{equation}
	Recently, black hole solutions have been constructed within this framework, in particular in regimes where the scalar field is spacelike ($\eta=-1$) and the multiplier is non-vanishing (${\bar \lambda}\neq0$)~\cite{Gorji:2020ten,Khodadi:2024ubi}. Closely related stealth scalar-hair constructions in higher-order scalar-tensor theories, including covariant GR-solution conditions and rotating Hamilton-Jacobi profiles, were studied in \cite{Charmousis:2019vnf}, \cite{Takahashi:2020hso}, \cite{Anson:2020trg} and \cite{BenAchour:2020fgy}.

	By varying the action with respect to $\lambda$, we find that the constraint equation \eqref{eq:constraint} reduces to the scalar mimetic constraint
	\begin{equation}\label{eq:mimetic-constraint}
		g^{\mu\nu}\partial_\mu\phi\,\partial_\nu\phi=-\eta,
	\end{equation}
	which implies that $\eta=+1$ corresponds to a timelike $\partial_\mu\phi$ and $\eta=-1$ to a spacelike one. In what follows, we will primarily focus on the timelike case ($\eta=+1$).
	
	After using the constraint \eqref{eq:mimetic-constraint}, the metric equations follow from the general form given in \eqref{eq:metric-eom-general-matter}. For the choice in
	\eqref{eq:C_scalar_choice} we find
	\begin{equation}\label{eq:EEs}
		G_{\mu\nu}=-2\lambda\,\partial_\mu\phi\,\partial_\nu\phi+T^{\rm m}_{\mu\nu}\,.
	\end{equation}
	This equation  indicates that the scalar field configuration acts as an effective energy-momentum tensor sourced by the Lagrange multiplier. For instance, in a Friedmann-Lema\^itre-Robertson-Walker (FLRW) universe, the scalar contribution on the r.h.s. behaves as a pressureless perfect fluid, with the effective energy density $-2\lambda$ and $\phi$ playing the role of a velocity potential. 
	
	The scalar field equation follows from the general Euler-Lagrange equation in \eqref{eq:psi-eom-general} and reads
	\begin{equation}\label{eq:SF}
		\nabla_\mu J^\mu=0 \,; \qquad J^\mu\equiv\lambda g^{\mu\nu}\partial_\nu\phi \,.
	\end{equation}
	
	We now focus on the stealth branch identified in Sec.~\ref{sec-1}, namely backgrounds of the form
	\begin{align}\label{eq:BG-config-scalar}
		\{
		\bar g_{\mu\nu},\;\bar\phi,\;\bar\lambda=0,\;\bar\Xi
		\}
		\qquad\text{with}\qquad
		{\bar C} = {\bar g}^{\mu\nu}\partial_\mu\bar{\phi}\,\partial_\nu{\bar \phi}+1=0.
	\end{align}
	On such backgrounds, the scalar equation \eqref{eq:SF} is automatically satisfied, while the metric equations \eqref{eq:EEs} reduce to the Einstein equations \eqref{eq:EE-matter} in the presence of matter fields. The only nontrivial requirement on the scalar profile is the background constraint  provided in \eqref{eq:mimetic-constraint}.
	
	To solve the scalar field profiles, we note that the constraint equation \eqref{eq:BG-config-scalar} coincides with a Hamilton-Jacobi equation for the scalar field. Since $\partial_\mu\bar\phi$ is timelike, one may interpret $\bar\phi$ as a Hamilton principal function generating a timelike geodesic congruence on the GR geometry $\bar g_{\mu\nu}$. Physically, choosing $\bar\phi$ as the time coordinate fixes the lapse function to unity, which after choosing vanishing shift vector by means of the spatial coordinates, coincides with the synchronous gauge.
	
	It is worth mentioning that since the timelike scalar profile defines a geodesic congruence, its smoothness is generally local and caustics may form in the usual mimetic/dust sense. Higher-derivative imperfect mimetic extensions can modify this congruence structure and have been proposed as a way to avoid caustics \cite{Gorji:2025ajb}.
	
	Therefore, in the scalar realization the general statement of Sec.~\ref{sec-1} becomes explicit: every GR metric (including black hole geometries) admits a family of stealth configurations labeled by scalar profiles $\bar\phi$ solving Hamilton-Jacobi equation \eqref{eq:BG-config-scalar}. These examples are summarized in Table~\ref{tab:staticfr}. The static and rotating Hamilton-Jacobi derivations are given in Appendix~\ref{app:profiles}.

	\begin{table}[h]
		\centering
		\renewcommand{\arraystretch}{1.3}
		\begin{tabular}{|c|c|c|}
			\hline
			{\bf Stealth solution} & ${\cal L}_{\rm m}$ & $f(r)$ or $\Delta(r)$ \\
			\hline
			Schwarzschild & $0$ & $1-\frac{2M}{r}$\\
			\hline
			Schwarzschild-de-Sitter & $-\Lambda$ & $1-\frac{2M}{r}-\frac{1}{3}\Lambda r^2$\\
			\hline
			Reissner-Nordstr\"om & $-\frac14 F_{\mu\nu}F^{\mu\nu}$ & $1-\frac{2M}{r}+\frac{Q^2}{r^2}$\\
			\hline
			Reissner-Nordstr\"om-de-Sitter & $-\frac14 F_{\mu\nu}F^{\mu\nu}-\Lambda$ & $1-\frac{2M}{r}+\frac{Q^2}{r^2}-\frac{1}{3}\Lambda r^2$\\
			\hline
			Kerr & $0$ & $r^2-2Mr+a^2$\\
			\hline
			Kerr-Newman & $-\frac14 F_{\mu\nu}F^{\mu\nu}$ & $r^2-2Mr+a^2+Q^2$\\
			\hline
		\end{tabular}
		\caption{Static and rotating stealth backgrounds. The first four rows use $f(r)$ in \eqref{eq:metric-Schw}, and the last two use $\Delta(r)$ in \eqref{eq:metric-Kerr}. Here $M$, $Q$, $a$, and $\Lambda$ denote the mass, electric charge, rotation parameter, and cosmological constant, respectively. The scalar profiles have the form \eqref{eq:phi-sep}, with radial and angular functions determined in Appendix~\ref{app:profiles}.}
		\label{tab:staticfr}
	\end{table}
	
		\section{Perturbations around the stealth solutions}\label{sec-pert}
	
	In the previous section we illustrated stealth profiles on rotating and non-rotating black hole backgrounds. These backgrounds are of particular interest since GR black holes are strongly supported by astrophysical observations, including gravitational wave detections \cite{LIGOScientific:2016aoc,LIGOScientific:2016lio,LIGOScientific:2020tif,LIGOScientific:2021sio} and horizon-scale imaging \cite{EventHorizonTelescope:2019dse,EventHorizonTelescope:2022wkp}. However, the stealth configurations coincide with the corresponding GR solutions only at the background level. To assess whether the mimetic sector can leave observable imprints, one must study fluctuations around these backgrounds. In this section we therefore analyze perturbations around the general family of stealth solutions with $\bar{\lambda}=0$ and examine how the perturbative description depends on whether $\bar\lambda$ vanishes exactly or is merely small.
	
	Before doing so, let us point out a simple obstruction that already appears at the level of the background equations in the scalar realization. Consider the static and spherically symmetric metric \eqref{eq:metric-Schw} together with the timelike Hamilton-Jacobi profile \eqref{eq:phi-sep}. For the standard static matter sources considered in Table~\ref{tab:staticfr}, one has $\bar T^{\rm m}_{tr}=0$, while the metric \eqref{eq:metric-Schw} gives $\bar G_{tr}=0$. Therefore, the $tr$ component of the scalar-field Einstein equation \eqref{eq:EEs} gives
	\begin{align}
		0=\bar G_{tr}
		= -2\bar\lambda\,\partial_t\bar\phi\,\partial_r\bar\phi
		=2E\bar\lambda\,{\mathrm R}'(r) \, .
	\end{align}
	Thus, for a generic timelike scalar profile with $E{\mathrm R}'(r)\neq0$, keeping the same static diagonal metric ansatz forces $\bar\lambda=0$. Equivalently, within this ansatz, the exactly stealth scalar background is not continuously connected to a background with the same Hamilton-Jacobi profile and nonzero $\bar\lambda(r)$. To keep $\bar\lambda\neq0$, one must either change the scalar profile, allow a non-diagonal or time-dependent geometry, or leave the static diagonal ansatz. This is a background-level obstruction specific to the scalar realization. We now turn to the more general perturbative question, which applies to arbitrary constraint functionals $C[g,\Psi]$ and does not rely on any particular background geometry.
	
	We expand all fields around a background solution as
	\begin{equation}\label{eq:pert-expansion}
		g_{\mu\nu}=\bar g_{\mu\nu}+h_{\mu\nu}\,,\qquad
		\lambda=\bar\lambda+\delta\lambda\,,\qquad
		\Psi=\bar\Psi+\delta\Psi\,, \qquad \Xi=\bar\Xi+\delta\Xi\,,
	\end{equation}
	and correspondingly expand the constraint functional,
	\begin{equation}\label{eq:C-expansion}
		C=\bar C + C^{(1)}+\frac12 C^{(2)}+\frac{1}{3!}C^{(3)}+\cdots,
	\end{equation}
	where $C^{(n)}$ denotes the part of $C$ that is $n$th order in the fluctuations $(h,\delta\Psi)$.
	
	It is also convenient to expand the metric determinant factor as
	\begin{equation}\label{eq:sqrtg-expansion}
		\sqrt{-g}=\sqrt{-\bar g}\left(
		1 + J^{(1)}+\frac12 J^{(2)}+\cdots
		\right),
	\end{equation}
	where $J^{(n)}$ is $n$th order in $h_{\mu\nu}$ (e.g.\ $J^{(1)}=\frac12 h$ with $h\equiv \bar g^{\mu\nu}h_{\mu\nu}$). Throughout this section we use the notation $(\cdots)^{(n)}$ for the $n$th-order piece in fluctuations.
	
	We split the action \eqref{eq:action} into a gravitational and matter part $S_{\rm GR}$ and the mimetic constraint part $S_{\rm c}$,
	\begin{equation}
		\begin{gathered}
			S=S_{\rm GR} + S_{\rm c} \,,
			\\
			S_{\rm GR}\equiv \int \D^4x\,\sqrt{-g}\left(\frac{1}{2}R+{\cal L}_{\rm m}[g,\Xi]\right),
			\quad
			S_{\rm c}\equiv \int \D^4x\,\sqrt{-g}\,\lambda\,C[g,\Psi].
		\end{gathered}
	\end{equation}
	The corresponding quadratic action is
	\begin{equation}
		S^{(2)}=S_{\rm GR}^{(2)}+S_{\rm c}^{(2)}.
	\end{equation}
	The piece $S_{\rm GR}^{(2)}$ is the standard quadratic action for the metric and matter perturbations around the background $\bar g_{\mu\nu}$, and we will not rewrite it explicitly. Since the matter fields couple to $g_{\mu\nu}$ only and not to $(\lambda,\Psi)$, the constraint sector $S_{\rm c}$ is insensitive to their presence. Consequently, every background GR solution sourced by such matter, not only the vacuum ones, is a stealth solution at $\bar\lambda=0$ whenever $\bar C=0$ can be satisfied, and the branch analysis below is unchanged.
	
	The nontrivial point is the mimetic constrained sector. Expanding the product $\sqrt{-g}\,\lambda\,C$ to second order yields
	\begin{align}\label{eq:Sm2_general}
		S_{\rm c}^{(2)}=\int \D^4x\,\sqrt{-\bar g}\Big[
		\frac{\bar\lambda}{2}\,C^{(2)}
		+\delta\lambda\,C^{(1)}
		+J^{(1)}\left(\bar\lambda\,C^{(1)}+\delta\lambda\,\bar C\right)
		+\frac12 J^{(2)}\,\bar\lambda\,\bar C
		\Big] \,.
	\end{align}
	On backgrounds that satisfy the constraint, $\bar C=0$, \eqref{eq:Sm2_general} simplifies to
	\begin{equation}\label{eq:Sm2_barC0}
		S_{\rm c}^{(2)}=\int \D^4x\,\sqrt{-\bar g}\Big[
		\frac{\bar\lambda}{2}\,C^{(2)}
		+\delta\lambda\,C^{(1)}
		+J^{(1)}\,\bar\lambda\,C^{(1)}
		\Big].
	\end{equation}
	Independently of the value of $\bar\lambda$, variation of the quadratic action with respect to $\delta\lambda$ gives the linearized constraint
	\begin{equation}\label{eq:linearized_constraint}
		C^{(1)}=0,
	\end{equation}
	which is the first nontrivial order in the perturbative expansion of the exact constraint $C=0$. At higher orders, this same equation relates higher-order field corrections to products of lower-order fluctuations. The structure \eqref{eq:Sm2_barC0} makes the branch dependence transparent. We thus compare $\bar\lambda\neq 0$ and $\bar\lambda=0$ in realizations admitting both branches.
	
	If $\bar\lambda\neq 0$, the value of the quadratic constraint action on \eqref{eq:linearized_constraint} is
	\begin{equation}\label{eq:Sm2_branchA}
		S_{\rm c}^{(2)}\Big|_{{\bar\lambda}\neq 0,\,C^{(1)}=0}
		=\int \D^4x\,\sqrt{-\bar g}\;\frac{\bar\lambda}{2}\,C^{(2)}.
	\end{equation}
	Although $\lambda$ enters the action linearly, the background value $\bar\lambda$ acts as an overall weight for the quadratic structures contained in $C^{(2)}$. For many choices of $C[g,\Psi]$, $C^{(2)}$ includes two-derivative terms for some combination(s) of $(\delta\Psi,h_{\mu\nu})$, which may contribute to the quadratic dynamics of an additional mode. In this sense the $\bar\lambda\neq 0$ branch is typically the generic one from the viewpoint of the perturbation analysis.
	
	If $\bar\lambda=0$, the same evaluation gives
	\begin{equation}\label{eq:Sm2_branchB_onshell}
		S_{\rm c}^{(2)}\Big|_{\bar\lambda=0,\,C^{(1)}=0}=0.
	\end{equation}
	Before imposing the constraint, the remaining term $\delta\lambda\,C^{(1)}$ must be retained when varying the quadratic action. It imposes $C^{(1)}=0$ and also contributes to the other linearized field equations. Thus the vanishing action value in \eqref{eq:Sm2_branchB_onshell} does not imply GR perturbation dynamics.
	
	Comparing \eqref{eq:Sm2_branchA} and \eqref{eq:Sm2_branchB_onshell}, the quadratic structures weighted by $\bar\lambda$ disappear on the stealth branch. If the $\bar\lambda$-weighted quadratic terms contribute to the kinetic operator of a physical mode after the constraints are resolved, their disappearance at $\bar\lambda=0$ can lead to a {\it non-uniform} perturbative limit and potentially to {\it strong coupling}. The physical spectrum must be determined from the full constrained system rather than from these restricted action values alone.
	
	Linearity in $\lambda$ does not remove nonlinear interactions in the fluctuations. In particular, the cubic action contains $\delta\lambda C^{(2)}/2$ even when $\bar\lambda=0$. A vanishing kinetic coefficient in the presence of interactions can signal strong coupling. Establishing the coupling scale requires canonical normalization after solving the constraints.

	Let us now illustrate the general analysis in an explicit realization: a scalar field on a fixed Minkowski metric. We freeze the metric perturbations in order to isolate the constraint sector acting on the scalar. For nonzero $\bar\lambda$, this is a fixed-metric example, not a vacuum solution of the coupled Einstein equations. Taking the
	scalar realization $\Psi=\phi$ with $C[g,\phi]=g^{\mu\nu}\partial_\mu\phi\,\partial_\nu\phi+\eta$ introduced
	above, we consider
	\begin{align}
		g_{\mu\nu}=\eta_{\mu\nu}\,, \qquad \phi=\bar\phi+\delta\phi\,, \qquad \lambda=\bar\lambda+\delta\lambda\,,
	\end{align}
	where $\bar\phi=\bar\phi(t)$ and $\bar\lambda$ is constant, on a background satisfying $\bar C=0$. With the
	convention of \eqref{eq:C-expansion}, the expansion of the constraint reads
	\begin{align}\label{eq:C-exp-scalar}
		C^{(1)}=2\,\eta^{\mu\nu}\bar\phi_\mu\,\delta\phi_\nu=-2\,\dot{\bar\phi}\,\delta\dot\phi\,,
		\qquad
		C^{(2)}=2\,\eta^{\mu\nu}\delta\phi_\mu\,\delta\phi_\nu
		=2\big[(\partial_i\delta\phi)^2-(\delta\dot\phi)^2\big]\,,
	\end{align}
	while $C^{(n)}=0$ for $n\ge3$ since $C$ is quadratic in $\phi$. The metric is frozen, so $J^{(n)}=0$ and
	the quadratic constraint Lagrangian \eqref{eq:Sm2_barC0} reduces to
	\begin{align}\label{eq:Lcon2_scalar}
		L_{\rm c}^{(2)}=\frac{\bar\lambda}{2}\,C^{(2)}+\delta\lambda\,C^{(1)}\,.
	\end{align}
	Variation with respect to $\delta\lambda$ reproduces the linearized constraint \eqref{eq:linearized_constraint},
	\begin{align}
		C^{(1)}=0 \quad\Longleftrightarrow\quad \dot{\bar\phi}\,\delta\dot\phi=0\,,
	\end{align}
	hence $\delta\dot\phi=0$ as long as $\dot{\bar\phi}\neq0$. On this surface the multiplier term in
	\eqref{eq:Lcon2_scalar} drops, and the constraint sector leaves
	\begin{align}\label{eq:Lcon2_scalar_os}
		L_{\rm c}^{(2)}\big|_{C^{(1)}=0}
		=\frac{\bar\lambda}{2}\,C^{(2)}\big|_{\delta\dot\phi=0}
		=\bar\lambda\,(\partial_i\delta\phi)^2\,,
	\end{align}
	which is the scalar example of \eqref{eq:Sm2_branchA}. Here the linearized constraint removes the time-kinetic part of $C^{(2)}$ and leaves a pure spatial gradient. This $\bar\lambda$-weighted term is present for every $\bar\lambda\neq0$ and vanishes at $\bar\lambda=0$. The multiplier term in \eqref{eq:Lcon2_scalar} must still be varied to obtain the scalar and multiplier equations. Therefore, the restricted gradient term alone does not determine the physical spectrum or establish weak coupling. Although $C^{(n)}=0$ for $n\ge3$ in this fixed-metric example, $\lambda C$ still contains the cubic interaction $\delta\lambda C^{(2)}/2$.

	If a particular realization has a degenerate physical perturbation operator on the stealth branch, a natural question is whether this can be cured by extending
	the constrained sector, and if so, at what cost. There are at least three qualitatively different possibilities; (i) One may add operators to the term $\lambda C$ such as $\nabla_\mu C\nabla^\mu C$, or more general functions $F(C,\nabla C,\ldots)$ that are at least quadratic in $C$ and its derivatives. This option is structurally conservative and, importantly, it preserves the stealth GR family: as long as
	the Lagrange-multiplier term $\lambda C$ is kept, $C=0$ remains an exact equation of motion, and on the constraint surface one has $C=0$ as well as $\nabla_\mu C=0$, $\Box C=0$, etc. These terms and their first variations vanish when $C=0$, so the GR backgrounds with $\bar\lambda=0$
	remain exact solutions. However, for the same reason these terms cannot resolve any such degeneracy, since they do not generate a genuine quadratic operator for the would-be extra mode on the stealth branch. (ii) A second possibility is to add invariants to $\lambda C$ that depend explicitly on $\nabla\Psi$ and/or $\nabla\nabla\Psi$, which cannot in general be rewritten as functions of $C$ and therefore can modify the perturbative
	dynamics. The price is that the stealth GR family is no longer automatic and must be checked in the deformed theory, and that such terms generically change the degree-of-freedom content (often introducing additional modes unless suitable degeneracy conditions are imposed).\footnote{For example, in the scalar realization $\Psi=\phi$ with $C[g,\phi]=g^{\mu\nu}\partial_\mu\phi\partial_\nu\phi+\eta$, the on-shell condition $\nabla_\mu C=0$ implies $2\partial^\alpha\phi\nabla_\mu\partial_\alpha\phi=0$. Defining $A^\mu{}_{\nu}\equiv\nabla^\mu\nabla_\nu\phi$, this implies that $\vartheta_n\equiv\partial_\mu\phi\,(A^n)^\mu{}_{\nu}\partial^\nu\phi$ vanishes for all $n\ge1$, where $A^n$ denotes the matrix power. Thus operators constructed solely from $\nabla_\mu C$ (or its contractions) cannot cure any such degeneracy, while genuinely new deformations may instead be built from invariants involving $\nabla\phi$
		and $\nabla\nabla\phi$, such as trace-type combinations $\chi_n\equiv\operatorname{tr}(A^n)$
		(in particular $\chi_1=\Box\phi$), which cannot be reexpressed as functions of the single scalar $C$
		\cite{Langlois:2018jdg}.} (iii) Finally, one may endow the multiplier with dynamics, e.g.\ by adding $(\nabla\lambda)^2$ (or kinetic
	mixing). This was first proposed in \cite{Golovnev:2013jxa}, though not directly in relation to the $\bar\lambda\to0$ limit in the specific scalar field realization. This modifies the quadratic dynamics and can introduce new degrees of freedom. A pure $(\nabla\lambda)^2$ addition preserves the existing $\bar\lambda=0$ stealth backgrounds, but away from this branch the $\lambda$ equation no longer reduces to the pointwise constraint $C=0$. In short,
	$C$-only deformations keep GR solutions but cannot cure any such perturbative degeneracy, while changing the physical perturbation dynamics requires additional terms whose effect on the spectrum and background solutions must be checked.

	\section{Summary and conclusions}\label{sec:summary}
	
	In this paper we studied a broad class of constrained mimetic-type extensions of GR defined by the action \eqref{eq:action}, where a Lagrange multiplier $\lambda$ enforces a constraint $C[g,\Psi]=0$ for generic field content $\Psi$. The functional $C[g,\Psi]$ is kept general and may depend on the metric, on $\Psi$, and on a finite number of covariant derivatives. This setup includes the standard scalar mimetic model \eqref{eq:C_scalar_choice}, as well as vector, gauge-field, and more general constrained realizations. We focused on local stealth GR solutions with $\bar\lambda=0$ and deviations from them.
	
	At the level of background solutions, we showed that the exactly stealth branch $\bar\lambda=0$ reproduces GR. On this branch, the constrained sector does not backreact on the metric, and any GR geometry is admitted as a local solution on domains where the background fields $\bar\Psi$ can be chosen to satisfy $\bar C=0$. We illustrated this explicitly in the scalar realization of Sec.~\ref{sec-scalar}, where the constraint becomes a Hamilton-Jacobi equation for the scalar profile and selects geodesic congruences on the underlying GR spacetime. For static and spherically symmetric backgrounds the equation is separable, while for rotating black holes Carter separability gives stealth Kerr profiles labeled by the conserved quantities of geodesic motion, as reviewed in Appendix~\ref{app:profiles}.
	
	The main result of the paper is that background agreement with GR does not by itself establish a regular perturbative limit. For nonzero $\bar\lambda$, the constrained sector contains quadratic structures weighted by $\bar\lambda$. On the exactly stealth branch these terms vanish, while the multiplier perturbation still enters the field equations and nonlinear interactions can remain. The constraint is enforced order by order as the expansion of the single equation $C=0$. If the vanishing quadratic terms contribute to the kinetic operator of a physical mode after the constraints are resolved, the $\bar\lambda\to0$ limit can be non-uniform and strong coupling can arise. Its detailed form depends on the choice of $C[g,\Psi]$ and the treatment of the full constrained system. Thus the perturbative behavior must be established separately from the existence of stealth GR backgrounds.
	
	\subsubsection*{Acknowledgements}
	The work of M.A.G. is supported by IBS under the project code, IBS-R018-D3.
	
	\appendix
	
	
	\section{Local deviations from Schwarzschild}
	\label{app:screenradius}
	
	In Eq.~\eqref{r-screen}, we estimated the relation between the local deviation scale $r_{\mathrm{dev}}$ and the background value of the Lagrange multiplier ${\bar\lambda}$ for a compact object of mass $M$, using the Kretschmann scalar to set the curvature scale. This gives a reasonable order-of-magnitude estimate. In this appendix, we recover the same scaling in the scalar field case by solving the equations to first order for a local, time-dependent deformation of Schwarzschild.
	
	For ${\bar \lambda}=0$, the Schwarzschild metric is an exact solution of the system. A nonzero ${\bar \lambda}$ induces deviations from this solution. To include the radial scalar flow consistently, we allow time dependence and use
	\begin{align}\label{app:metric}
		\D s^2=-\D t^2+\bigl(\D r+v(t,r)\D t\bigr)^2+r^2\D\Omega^2 \,;
		\qquad \bar\phi=-t \,.
	\end{align}
	The scalar constraint is satisfied identically. Writing $v^2=2m(t,r)/r$ with $v>0$, the Einstein equations give
	\begin{align}\label{app:mass}
		\partial_t m-v\partial_r m=0 \,; \qquad
		\bar\lambda=-\frac{\partial_r m}{r^2} \,.
	\end{align}
	The scalar equation \eqref{eq:SF} then follows from these relations. Schwarzschild corresponds to $m=M$. We set
	\begin{align}\label{app:dm}
		m=M+\delta m \,; \qquad |\delta m|\ll M \,.
	\end{align}
	To first order, the mass equation becomes
	\begin{align}\label{app:transport}
		\partial_t\delta m-\sqrt{\frac{2M}{r}}\,\partial_r\delta m=0 \,.
	\end{align}
	A local solution, with $C_0$ a small integration constant, is
	\begin{align}\label{app:lambda}
		\delta m=-C_0t-\frac{2C_0}{3\sqrt{2M}}r^{3/2} \,; \qquad
		\bar\lambda(r)=\frac{C_0}{\sqrt{2M}}r^{-3/2} \,.
	\end{align}
	Both expressions are accurate to first order in $C_0$. On any fixed regular domain where $|\delta m|\ll M$, this gives a parametric approach to the stealth solution as $C_0\to0$.
	
	Define $f(t,r)\equiv g^{rr}=1-2m(t,r)/r$. At the reference time $t=0$, we obtain
	\begin{align}\label{app-f}
		f(0,r)=1-\frac{2M}{r}+\delta f(r) \,; \qquad
		\delta f(r)=\frac{4C_0}{3\sqrt{2M}}r^{1/2} \,.
	\end{align}
	The sign of $C_0$ fixes the sign of $\bar\lambda$ and of the induced correction. Comparing its magnitude with the Schwarzschild term,
	\begin{align}\label{app-screen-condition}
		|\delta f(r_{\mathrm{dev}})|\sim\frac{2M}{r_{\mathrm{dev}}} \, ,
	\end{align}
	gives the order-of-magnitude estimate
	\begin{align}
		r_{\mathrm{dev}}\sim\frac{M}{|C_0|^{2/3}}
		\sim\left(\frac{M}{|\bar\lambda(r_{\mathrm{dev}})|}\right)^{1/3} \,.
	\end{align}
	This recovers the scaling in \eqref{r-screen} as an estimate of where the local deformation ceases to be small relative to the Schwarzschild term.
	
	\section{Scalar profiles on black hole backgrounds}\label{app:profiles}
	
	\subsection{Stealth static and spherical solutions}\label{sec:static}
	Let us begin with a static and spherical metric, which describes the spacetime outside a non-rotating compact object. 
	In standard coordinates, the line element is
	\begin{equation}\label{eq:metric-Schw}
		\D{s}^2=- f(r) \D t^2+ f(r)^{-1} \D r^2+r^2 \D\Omega^2 \,,
	\end{equation}
	where $\D\Omega^2$ is the metric of a unit sphere and for simplicity, we focus on the single-function case. Depending on the matter Lagrangian, the corresponding solutions for $f(r)$ are summarized in Table~\ref{tab:staticfr}.

	We seek a background configuration of the form \eqref{eq:BG-config-matter} with $\bar{\lambda}=0$. On this branch, the scalar contribution to the effective stress tensor vanishes, $\bar T^{\rm c}_{\mu\nu}=0$, and therefore the scalar profile does not backreact on the background geometry. For this reason, no symmetry assumption has to be imposed on $\bar\phi$, and we may start from the most general ansatz
	\begin{equation}\label{eq:BG-config-Schw}
		\bar{\phi}=\bar{\phi}(t,r,\theta,\varphi) \,.
	\end{equation}

	Substituting \eqref{eq:metric-Schw} and \eqref{eq:BG-config-Schw} in \eqref{eq:BG-config-scalar} we find
	\begin{equation}\label{eq:HJ-Schw}
		- \frac{1}{f(r)} \left(\frac{\partial{\bar \phi}}{\partial{t}}\right)^2
		+ f(r) \left(\frac{\partial{\bar \phi}}{\partial{r}}\right)^2 
		+ \frac{1}{r^2} \left(\frac{\partial{\bar \phi}}{\partial{\theta}}\right)^2 
		+ \frac{1}{r^2\sin^2\theta} \left(\frac{\partial{\bar \phi}}{\partial{\varphi}}\right)^2 = -1 \,.
	\end{equation}
	
	This equation governs the scalar field on the fixed background.
	A complete separated integral adapted to the Killing coordinates may be chosen with linear dependence on $t$ and $\varphi$. We therefore consider the ansatz
	\begin{align}\label{eq:phi-sep}
		{\bar \phi} = - E\,t + {\mathrm R}(r) + \Theta(\theta) + L\,\varphi \,, 
	\end{align}
	where $E$ and $L$ are constants. Substituting the above ansatz in (\ref{eq:HJ-Schw}), we obtain \begin{align}
		r^2 f(r) \left(\frac{\D{{\mathrm R}}}{\D{r}}\right)^2 
		- \frac{E^2 r^2}{f(r)}
		+ r^2 = - \left(\frac{\D\Theta}{\D\theta}\right)^2 - \frac{L^2}{\sin^2\theta} \,.
	\end{align}
	As the l.h.s. is only function of $r$ while the r.h.s. is only function of $\theta$, the only way to satisfy the above equation is that both sides are constant. Since the r.h.s. is non-positive, we consider
	\begin{align}
		\label{thetaSch}
		\left(\frac{\D\Theta}{\D\theta}\right)^2 + \frac{L^2}{\sin^2\theta} - K^2 = 0 \,,
	\end{align}
	where $K$ is a constant. For the radial part we need to solve
	\begin{align}
		f(r) \left(\frac{\D{{\mathrm R}}}{\D{r}}\right)^2 - \frac{E^2}{f(r)} + 1 + \frac{K^2}{r^2} = 0 \,.
	\end{align}
	
	The last two equations determine the angular and radial parts of the scalar profile by direct integration. They are precisely the separated Hamilton-Jacobi equations for a unit-mass timelike geodesic on the fixed static and spherically symmetric background, with $E$, $L$, and $K$ playing the role of separation constants. Since the theory is shift symmetric, only the gradient $\nabla_\mu\bar\phi$ is physically relevant for the background equations. The angular equation \eqref{thetaSch} is independent of the function $f(r)$ and gives the allowed range of polar motion; in particular, for $L\neq0$ the poles $\theta=0,\pi$ are excluded by the divergence of the $L^2/\sin^2\theta$ term. The radial part depends on the specific background through $f(r)$ and must be treated with care near horizons, where the static coordinates may become singular. This issue is coordinate dependent and can be analyzed using horizon-regular coordinates, while it does not affect the asymptotic behavior of the profile \cite{Chandrasekhar1983,MisnerThorneWheeler1973}.

	\subsection{Stealth rotating solutions}
	Let us now turn to the more general case of a rotating black hole, described by the Kerr-Newman metric. We work in Boyer-Lindquist coordinates, which are well suited to separating variables in rotating spacetimes. The metric is given by
	\begin{align}\label{eq:metric-Kerr}
		\begin{split}
			\D{s}^2 &=-\left(\frac{\Delta-a^2\sin^2\theta}{\rho^2}\right){\D}t^2
			-\frac{2a\sin^2\theta (r^2+a^2-\Delta)}{\rho^2} {\D}t{\D}\varphi
			+\frac{\rho^2}{\Delta}dr^2+\rho^2\D\theta^2
			\\
			&\hspace{3cm}
			+ \left(\frac{(r^2+a^2)^2 - \Delta a^2\sin^2\theta}{\rho^2} \right)
			\sin^2\theta \D\varphi^2 \,,
		\end{split}
	\end{align}
	where
	\begin{equation}
		\Delta(r)=r^2-2Mr+a^2+Q^2 \,,
		\qquad
		\rho^2(r,\theta)=r^2+a^2\cos^2\theta \,.
	\end{equation}
	The above metric reduces to Schwarzschild metric for $a=0$ and $Q=0$.
	
	To find ${\bar \phi}$, a complete separated integral adapted to the Killing coordinates may again be chosen with linear dependence on $t$ and $\varphi$, so we use the same ansatz as in \eqref{eq:phi-sep}.
	
	Substituting \eqref{eq:metric-Kerr} and \eqref{eq:phi-sep} in \eqref{eq:BG-config-scalar},  after some algebraic manipulation one can show that it takes the following form
	\begin{equation}\label{eq:HJ-separated-form}
		\begin{split}
			&\Delta(r) \left(\frac{\D{{\mathrm R}}}{\D{r}}\right)^2
			- \frac{A(r)^2}{\Delta(r)}
			+ r^2
			=
			- \left(\frac{\D{\Theta}}{\D{\theta}}\right)^2
			- \frac{\bigl(L - aE\sin^2\theta\bigr)^2}{\sin^2\theta}
			- a^2 \cos^2\theta \,,
		\end{split}
	\end{equation}
	where we have defined
	\begin{align}
		A(r) = \left(r^2+a^2\right) E - aL \,.
	\end{align}
	Since the r.h.s. of \eqref{eq:HJ-separated-form} depends only on $\theta$ and is non-positive from its explicit form, the equation can be separated by introducing a separation constant, as in the static case. The angular equation is then given by
	\begin{align}
		\left(\frac{\D{\Theta}}{\D{\theta}}\right)^2
		+ \frac{\bigl(L - aE\sin^2\theta\bigr)^2}{\sin^2\theta}
		+ a^2 \cos^2\theta - K^2 = 0 \,,
	\end{align}
	while the radial equation is
	\begin{align}
		\Delta(r) \left(\frac{\D{{\mathrm R}}}{\D{r}}\right)^2
		- \frac{A(r)^2}{\Delta(r)}
		+ r^2 + K^2 = 0 \,,
	\end{align}
	Here $K^2$ denotes the separation constant which is related to the conventional Carter constant by the usual shift involving $E$ and $L$. Asymptotically, $\D{R(r)}/{\D{r}}$ has the same leading behavior as in the static and spherically symmetric case. Explicit expressions for the radial and angular functions $R(r)$ and $\Theta(\theta)$ can be found in \cite{Chandrasekhar1983,MisnerThorneWheeler1973} for Kerr spacetime and in \cite{HackmannXu2013,Carter1968} for Kerr-Newman spacetime.
	
	\bibliographystyle{JHEPmod}
	\bibliography{refs}

\providecommand{\href}[2]{#2}\begingroup\raggedright\begin{thebibliography}{10}

\bibitem{Jain:2010ka}
B.~Jain and J.~Khoury, {\it {Cosmological Tests of Gravity}},
  \href{https://doi.org/10.1016/j.aop.2010.04.002}{Annals Phys. {\bfseries 325}
  (2010) 1479} [\href{http://arxiv.org/abs/1004.3294}{{\ttfamily
  arXiv:1004.3294}}].

\bibitem{Brax:2013ida}
P.~Brax, {\it {Screening mechanisms in modified gravity}},
  \href{https://doi.org/10.1088/0264-9381/30/21/214005}{Class. Quant. Grav.
  {\bfseries 30} (2013) 214005}.

\bibitem{Kobayashi:2014ida}
T.~Kobayashi, Y.~Watanabe and D.~Yamauchi, {\it {Breaking of Vainshtein
  screening in scalar-tensor theories beyond Horndeski}},
  \href{https://doi.org/10.1103/PhysRevD.91.064013}{Phys. Rev. D {\bfseries 91}
  (2015) 064013} [\href{http://arxiv.org/abs/1411.4130}{{\ttfamily
  arXiv:1411.4130}}].

\bibitem{Joyce:2014kja}
A.~Joyce, B.~Jain, J.~Khoury and M.~Trodden, {\it {Beyond the Cosmological
  Standard Model}}, \href{https://doi.org/10.1016/j.physrep.2014.12.002}{Phys.
  Rept. {\bfseries 568} (2015) 1}
  [\href{http://arxiv.org/abs/1407.0059}{{\ttfamily arXiv:1407.0059}}].

\bibitem{Koyama:2015vza}
K.~Koyama, {\it {Cosmological Tests of Modified Gravity}},
  \href{https://doi.org/10.1088/0034-4885/79/4/046902}{Rept. Prog. Phys.
  {\bfseries 79} (2016) 046902}
  [\href{http://arxiv.org/abs/1504.04623}{{\ttfamily arXiv:1504.04623}}].

\bibitem{Joyce:2016vqv}
A.~Joyce, L.~Lombriser and F.~Schmidt, {\it {Dark Energy Versus Modified
  Gravity}}, \href{https://doi.org/10.1146/annurev-nucl-102115-044553}{Ann.
  Rev. Nucl. Part. Sci. {\bfseries 66} (2016) 95}
  [\href{http://arxiv.org/abs/1601.06133}{{\ttfamily arXiv:1601.06133}}].

\bibitem{Burrage:2017qrf}
C.~Burrage and J.~Sakstein, {\it {Tests of Chameleon Gravity}},
  \href{https://doi.org/10.1007/s41114-018-0011-x}{Living Rev. Rel. {\bfseries
  21} (2018) 1} [\href{http://arxiv.org/abs/1709.09071}{{\ttfamily
  arXiv:1709.09071}}].

\bibitem{Crisostomi:2017lbg}
M.~Crisostomi and K.~Koyama, {\it {Vainshtein mechanism after GW170817}},
  \href{https://doi.org/10.1103/PhysRevD.97.021301}{Phys. Rev. D {\bfseries 97}
  (2018) 021301} [\href{http://arxiv.org/abs/1711.06661}{{\ttfamily
  arXiv:1711.06661}}].

\bibitem{Dima:2017pwp}
A.~Dima and F.~Vernizzi, {\it {Vainshtein Screening in Scalar-Tensor Theories
  before and after GW170817: Constraints on Theories beyond Horndeski}},
  \href{https://doi.org/10.1103/PhysRevD.97.101302}{Phys. Rev. D {\bfseries 97}
  (2018) 101302} [\href{http://arxiv.org/abs/1712.04731}{{\ttfamily
  arXiv:1712.04731}}].

\bibitem{Ishak:2018his}
M.~Ishak, {\it {Testing General Relativity in Cosmology}},
  \href{https://doi.org/10.1007/s41114-018-0017-4}{Living Rev. Rel. {\bfseries
  22} (2019) 1} [\href{http://arxiv.org/abs/1806.10122}{{\ttfamily
  arXiv:1806.10122}}].

\bibitem{Amendola:2019laa}
L.~Amendola, D.~Bettoni, A.~M. Pinho and S.~Casas, {\it {Measuring gravity at
  cosmological scales}},
  \href{https://doi.org/10.3390/universe6020020}{Universe {\bfseries 6} (2020)
  20} [\href{http://arxiv.org/abs/1902.06978}{{\ttfamily arXiv:1902.06978}}].

\bibitem{Kobayashi:2019hrl}
T.~Kobayashi, {\it {Horndeski theory and beyond: a review}},
  \href{https://doi.org/10.1088/1361-6633/ab2429}{Rept. Prog. Phys. {\bfseries
  82} (2019) 086901} [\href{http://arxiv.org/abs/1901.07183}{{\ttfamily
  arXiv:1901.07183}}].

\bibitem{Ferreira:2019xrr}
P.~G. Ferreira, {\it {Cosmological Tests of Gravity}},
  \href{https://doi.org/10.1146/annurev-astro-091918-104423}{Ann. Rev. Astron.
  Astrophys. {\bfseries 57} (2019) 335}
  [\href{http://arxiv.org/abs/1902.10503}{{\ttfamily arXiv:1902.10503}}].

\bibitem{Sakstein:2018fwz}
J.~Sakstein, {\it {Astrophysical tests of screened modified gravity}},
  \href{https://doi.org/10.1142/S0218271818480085}{Int. J. Mod. Phys. D
  {\bfseries 27} (2018) 1848008}
  [\href{http://arxiv.org/abs/2002.04194}{{\ttfamily arXiv:2002.04194}}].

\bibitem{Brax:2021wcv}
P.~Brax, S.~Casas, H.~Desmond and B.~Elder, {\it {Testing Screened Modified
  Gravity}}, \href{https://doi.org/10.3390/universe8010011}{Universe {\bfseries
  8} (2021) 11} [\href{http://arxiv.org/abs/2201.10817}{{\ttfamily
  arXiv:2201.10817}}].

\bibitem{Fischer:2024eic}
H.~Fischer, C.~K{\"a}ding and M.~Pitschmann, {\it {Screened Scalar Fields in
  the Laboratory and the Solar System}},
  \href{https://doi.org/10.3390/universe10070297}{Universe {\bfseries 10}
  (2024) 297} [\href{http://arxiv.org/abs/2405.14638}{{\ttfamily
  arXiv:2405.14638}}].

\bibitem{Kimura:2011dc}
R.~Kimura, T.~Kobayashi and K.~Yamamoto, {\it {Vainshtein screening in a
  cosmological background in the most general second-order scalar-tensor
  theory}}, \href{https://doi.org/10.1103/PhysRevD.85.024023}{Phys. Rev. D
  {\bfseries 85} (2012) 024023}
  [\href{http://arxiv.org/abs/1111.6749}{{\ttfamily arXiv:1111.6749}}].

\bibitem{Chamseddine:2013kea}
A.~H. Chamseddine and V.~Mukhanov, {\it {Mimetic Dark Matter}},
  \href{https://doi.org/10.1007/JHEP11(2013)135}{JHEP {\bfseries 11} (2013)
  135} [\href{http://arxiv.org/abs/1308.5410}{{\ttfamily arXiv:1308.5410}}].

\bibitem{Deruelle:2014zza}
N.~Deruelle and J.~Rua, {\it {Disformal Transformations, Veiled General
  Relativity and Mimetic Gravity}},
  \href{https://doi.org/10.1088/1475-7516/2014/09/002}{JCAP {\bfseries 09}
  (2014) 002} [\href{http://arxiv.org/abs/1407.0825}{{\ttfamily
  arXiv:1407.0825}}].

\bibitem{Arroja:2015wpa}
F.~Arroja, N.~Bartolo, P.~Karmakar and S.~Matarrese, {\it {The two faces of
  mimetic Horndeski gravity: disformal transformations and Lagrange
  multiplier}}, \href{https://doi.org/10.1088/1475-7516/2015/09/051}{JCAP
  {\bfseries 09} (2015) 051} [\href{http://arxiv.org/abs/1506.08575}{{\ttfamily
  arXiv:1506.08575}}].

\bibitem{Golovnev:2013jxa}
A.~Golovnev, {\it {On the recently proposed Mimetic Dark Matter}},
  \href{https://doi.org/10.1016/j.physletb.2013.11.026}{Phys. Lett. B
  {\bfseries 728} (2014) 39} [\href{http://arxiv.org/abs/1310.2790}{{\ttfamily
  arXiv:1310.2790}}].

\bibitem{Barvinsky:2013mea}
A.~O. Barvinsky, {\it {Dark matter as a ghost free conformal extension of
  Einstein theory}}, \href{https://doi.org/10.1088/1475-7516/2014/01/014}{JCAP
  {\bfseries 01} (2014) 014} [\href{http://arxiv.org/abs/1311.3111}{{\ttfamily
  arXiv:1311.3111}}].

\bibitem{Chamseddine:2014vna}
A.~H. Chamseddine, V.~Mukhanov and A.~Vikman, {\it {Cosmology with Mimetic
  Matter}}, \href{https://doi.org/10.1088/1475-7516/2014/06/017}{JCAP
  {\bfseries 06} (2014) 017} [\href{http://arxiv.org/abs/1403.3961}{{\ttfamily
  arXiv:1403.3961}}].

\bibitem{Chaichian:2014qba}
M.~Chaichian, J.~Kluson, M.~Oksanen and A.~Tureanu, {\it {Mimetic dark matter,
  ghost instability and a mimetic tensor-vector-scalar gravity}},
  \href{https://doi.org/10.1007/JHEP12(2014)102}{JHEP {\bfseries 12} (2014)
  102} [\href{http://arxiv.org/abs/1404.4008}{{\ttfamily arXiv:1404.4008}}].

\bibitem{Mirzagholi:2014ifa}
L.~Mirzagholi and A.~Vikman, {\it {Imperfect Dark Matter}},
  \href{https://doi.org/10.1088/1475-7516/2015/06/028}{JCAP {\bfseries 06}
  (2015) 028} [\href{http://arxiv.org/abs/1412.7136}{{\ttfamily
  arXiv:1412.7136}}].

\bibitem{BenAchour:2016cay}
J.~Ben~Achour, D.~Langlois and K.~Noui, {\it {Degenerate higher order
  scalar-tensor theories beyond Horndeski and disformal transformations}},
  \href{https://doi.org/10.1103/PhysRevD.93.124005}{Phys. Rev. D {\bfseries 93}
  (2016) 124005} [\href{http://arxiv.org/abs/1602.08398}{{\ttfamily
  arXiv:1602.08398}}].

\bibitem{Sebastiani:2016ras}
L.~Sebastiani, S.~Vagnozzi and R.~Myrzakulov, {\it {Mimetic gravity: a review
  of recent developments and applications to cosmology and astrophysics}},
  \href{https://doi.org/10.1155/2017/3156915}{Adv. High Energy Phys. {\bfseries
  2017} (2017) 3156915} [\href{http://arxiv.org/abs/1612.08661}{{\ttfamily
  arXiv:1612.08661}}].

\bibitem{Firouzjahi:2017txv}
H.~Firouzjahi, M.~A. Gorji and S.~A. Hosseini~Mansoori, {\it {Instabilities in
  Mimetic Matter Perturbations}},
  \href{https://doi.org/10.1088/1475-7516/2017/07/031}{JCAP {\bfseries 07}
  (2017) 031} [\href{http://arxiv.org/abs/1703.02923}{{\ttfamily
  arXiv:1703.02923}}].

\bibitem{Zheng:2017qfs}
Y.~Zheng, L.~Shen, Y.~Mou and M.~Li, {\it {On (in)stabilities of perturbations
  in mimetic models with higher derivatives}},
  \href{https://doi.org/10.1088/1475-7516/2017/08/040}{JCAP {\bfseries 08}
  (2017) 040} [\href{http://arxiv.org/abs/1704.06834}{{\ttfamily
  arXiv:1704.06834}}].

\bibitem{Takahashi:2017pje}
K.~Takahashi and T.~Kobayashi, {\it {Extended mimetic gravity: Hamiltonian
  analysis and gradient instabilities}},
  \href{https://doi.org/10.1088/1475-7516/2017/11/038}{JCAP {\bfseries 11}
  (2017) 038} [\href{http://arxiv.org/abs/1708.02951}{{\ttfamily
  arXiv:1708.02951}}].

\bibitem{Hirano:2017zox}
S.~Hirano, S.~Nishi and T.~Kobayashi, {\it {Healthy imperfect dark matter from
  effective theory of mimetic cosmological perturbations}},
  \href{https://doi.org/10.1088/1475-7516/2017/07/009}{JCAP {\bfseries 07}
  (2017) 009} [\href{http://arxiv.org/abs/1704.06031}{{\ttfamily
  arXiv:1704.06031}}].

\bibitem{Gorji:2017cai}
M.~A. Gorji, S.~A. Hosseini~Mansoori and H.~Firouzjahi, {\it {Higher Derivative
  Mimetic Gravity}}, \href{https://doi.org/10.1088/1475-7516/2018/01/020}{JCAP
  {\bfseries 01} (2018) 020} [\href{http://arxiv.org/abs/1709.09988}{{\ttfamily
  arXiv:1709.09988}}].

\bibitem{Vagnozzi:2017ilo}
S.~Vagnozzi, {\it {Recovering a MOND-like acceleration law in mimetic
  gravity}}, \href{https://doi.org/10.1088/1361-6382/aa838b}{Class. Quant.
  Grav. {\bfseries 34} (2017) 185006}
  [\href{http://arxiv.org/abs/1708.00603}{{\ttfamily arXiv:1708.00603}}].

\bibitem{Langlois:2018jdg}
D.~Langlois, M.~Mancarella, K.~Noui and F.~Vernizzi, {\it {Mimetic gravity as
  DHOST theories}}, \href{https://doi.org/10.1088/1475-7516/2019/02/036}{JCAP
  {\bfseries 02} (2019) 036} [\href{http://arxiv.org/abs/1802.03394}{{\ttfamily
  arXiv:1802.03394}}].

\bibitem{Firouzjahi:2018xob}
H.~Firouzjahi, M.~A. Gorji, S.~A. Hosseini~Mansoori, A.~Karami and T.~Rostami,
  {\it {Two-field disformal transformation and mimetic cosmology}},
  \href{https://doi.org/10.1088/1475-7516/2018/11/046}{JCAP {\bfseries 11}
  (2018) 046} [\href{http://arxiv.org/abs/1806.11472}{{\ttfamily
  arXiv:1806.11472}}].

\bibitem{Gorji:2018okn}
M.~A. Gorji, S.~Mukohyama, H.~Firouzjahi and S.~A. Hosseini~Mansoori, {\it
  {Gauge Field Mimetic Cosmology}},
  \href{https://doi.org/10.1088/1475-7516/2018/08/047}{JCAP {\bfseries 08}
  (2018) 047} [\href{http://arxiv.org/abs/1807.06335}{{\ttfamily
  arXiv:1807.06335}}].

\bibitem{Jirousek:2018ago}
P.~Jirou{\v{s}}ek and A.~Vikman, {\it {New Weyl-invariant vector-tensor theory
  for the cosmological constant}},
  \href{https://doi.org/10.1088/1475-7516/2019/04/004}{JCAP {\bfseries 04}
  (2019) 004} [\href{http://arxiv.org/abs/1811.09547}{{\ttfamily
  arXiv:1811.09547}}].

\bibitem{Casalino:2018tcd}
A.~Casalino, M.~Rinaldi, L.~Sebastiani and S.~Vagnozzi, {\it {Mimicking dark
  matter and dark energy in a mimetic model compatible with GW170817}},
  \href{https://doi.org/10.1016/j.dark.2018.10.001}{Phys. Dark Univ. {\bfseries
  22} (2018) 108} [\href{http://arxiv.org/abs/1803.02620}{{\ttfamily
  arXiv:1803.02620}}].

\bibitem{Gorji:2019ttx}
M.~A. Gorji, S.~Mukohyama and H.~Firouzjahi, {\it {Cosmology in Mimetic SU(2)
  Gauge Theory}}, \href{https://doi.org/10.1088/1475-7516/2019/05/019}{JCAP
  {\bfseries 05} (2019) 019} [\href{http://arxiv.org/abs/1903.04845}{{\ttfamily
  arXiv:1903.04845}}].

\bibitem{Ganz:2019vre}
A.~Ganz, N.~Bartolo and S.~Matarrese, {\it {Towards a viable effective field
  theory of mimetic gravity}},
  \href{https://doi.org/10.1088/1475-7516/2019/12/037}{JCAP {\bfseries 12}
  (2019) 037} [\href{http://arxiv.org/abs/1907.10301}{{\ttfamily
  arXiv:1907.10301}}].

\bibitem{Gorji:2020ten}
M.~A. Gorji, A.~Allahyari, M.~Khodadi and H.~Firouzjahi, {\it {Mimetic black
  holes}}, \href{https://doi.org/10.1103/PhysRevD.101.124060}{Phys. Rev. D
  {\bfseries 101} (2020) 124060}
  [\href{http://arxiv.org/abs/1912.04636}{{\ttfamily arXiv:1912.04636}}].

\bibitem{Mansoori:2021fjd}
S.~A.~H. Mansoori, A.~Talebian, Z.~Molaee and H.~Firouzjahi, {\it {Multifield
  mimetic gravity}}, \href{https://doi.org/10.1103/PhysRevD.105.023529}{Phys.
  Rev. D {\bfseries 105} (2022) 023529}
  [\href{http://arxiv.org/abs/2108.11666}{{\ttfamily arXiv:2108.11666}}].

\bibitem{Jirousek:2022rym}
P.~Jirou{\v{s}}ek, K.~Shimada, A.~Vikman and M.~Yamaguchi, {\it {Disforming to
  conformal symmetry}},
  \href{https://doi.org/10.1088/1475-7516/2022/11/019}{JCAP {\bfseries 11}
  (2022) 019} [\href{http://arxiv.org/abs/2207.12611}{{\ttfamily
  arXiv:2207.12611}}].

\bibitem{Zheng:2022vwm}
Y.~Zheng and H.~Rao, {\it {Extensions of two-field mimetic gravity}},
  \href{https://doi.org/10.1007/JHEP04(2023)042}{JHEP {\bfseries 04} (2023)
  042} [\href{http://arxiv.org/abs/2210.10499}{{\ttfamily arXiv:2210.10499}}].

\bibitem{Jirousek:2022kli}
P.~Jirou{\v{s}}ek, K.~Shimada, A.~Vikman and M.~Yamaguchi, {\it {Mimetic
  K-essence}},  \href{http://arxiv.org/abs/2212.14867}{{\ttfamily
  arXiv:2212.14867}}.

\bibitem{Domenech:2023ryc}
G.~Dom{\`e}nech and A.~Ganz, {\it {Disformal symmetry in the Universe: mimetic
  gravity and beyond}},
  \href{https://doi.org/10.1088/1475-7516/2023/08/046}{JCAP {\bfseries 08}
  (2023) 046} [\href{http://arxiv.org/abs/2304.11035}{{\ttfamily
  arXiv:2304.11035}}].

\bibitem{Visa:2024fii}
D.-I. Visa, T.~Harko and S.~Shahidi, {\it {Mimetic Weyl geometric gravity}},
  \href{https://doi.org/10.1016/j.dark.2024.101720}{Phys. Dark Univ. {\bfseries
  46} (2024) 101720} [\href{http://arxiv.org/abs/2410.22787}{{\ttfamily
  arXiv:2410.22787}}].

\bibitem{Colleaux:2025vtm}
A.~Coll{\'e}aux and K.~Noui, {\it {Degenerate higher-order Maxwell-Einstein
  theories}}, \href{https://doi.org/10.1007/JHEP10(2025)007}{JHEP {\bfseries
  10} (2025) 007} [\href{http://arxiv.org/abs/2502.03311}{{\ttfamily
  arXiv:2502.03311}}].

\bibitem{Gorji:2025ajb}
M.~A. Gorji, {\it {Imperfect dark matter with higher derivatives}},
  \href{http://arxiv.org/abs/2510.23838}{{\ttfamily arXiv:2510.23838}}.

\bibitem{Gorji:2025fbv}
M.~A. Gorji, S.~Jana and P.~Petrov, {\it {Abelian and non-Abelian mimetic black
  holes}},  \href{http://arxiv.org/abs/2511.22062}{{\ttfamily
  arXiv:2511.22062}}.

\bibitem{Jacobson:2000xp}
T.~Jacobson and D.~Mattingly, {\it {Gravity with a dynamical preferred frame}},
  \href{https://doi.org/10.1103/PhysRevD.64.024028}{Phys. Rev. D {\bfseries 64}
  (2001) 024028} [\href{http://arxiv.org/abs/gr-qc/0007031}{{\ttfamily
  arXiv:gr-qc/0007031}}].

\bibitem{Blas:2009yd}
D.~Blas, O.~Pujolas and S.~Sibiryakov, {\it {On the Extra Mode and
  Inconsistency of Horava Gravity}},
  \href{https://doi.org/10.1088/1126-6708/2009/10/029}{JHEP {\bfseries 10}
  (2009) 029} [\href{http://arxiv.org/abs/0906.3046}{{\ttfamily
  arXiv:0906.3046}}].

\bibitem{Hammer:2020dqp}
K.~Hammer, P.~Jirousek and A.~Vikman, {\it {Axionic cosmological constant}},
  \href{http://arxiv.org/abs/2001.03169}{{\ttfamily arXiv:2001.03169}}.

\bibitem{Khodadi:2024ubi}
M.~Khodadi, S.~Vagnozzi and J.~T. Firouzjaee, {\it {Event Horizon Telescope
  observations exclude compact objects in baseline mimetic gravity}},
  \href{https://doi.org/10.1038/s41598-024-78264-y}{Sci. Rep. {\bfseries 14}
  (2024) 26932} [\href{http://arxiv.org/abs/2408.03241}{{\ttfamily
  arXiv:2408.03241}}].

\bibitem{Charmousis:2019vnf}
C.~Charmousis, M.~Crisostomi, R.~Gregory and N.~Stergioulas, {\it {Rotating
  Black Holes in Higher Order Gravity}},
  \href{https://doi.org/10.1103/PhysRevD.100.084020}{Phys. Rev. D {\bfseries
  100} (2019) 084020} [\href{http://arxiv.org/abs/1903.05519}{{\ttfamily
  arXiv:1903.05519}}].

\bibitem{Takahashi:2020hso}
K.~Takahashi and H.~Motohashi, {\it {General Relativity solutions with stealth
  scalar hair in quadratic higher-order scalar-tensor theories}},
  \href{https://doi.org/10.1088/1475-7516/2020/06/034}{JCAP {\bfseries 06}
  (2020) 034} [\href{http://arxiv.org/abs/2004.03883}{{\ttfamily
  arXiv:2004.03883}}].

\bibitem{Anson:2020trg}
T.~Anson, E.~Babichev, C.~Charmousis and M.~Hassaine, {\it {Disforming the Kerr
  metric}}, \href{https://doi.org/10.1007/JHEP01(2021)018}{JHEP {\bfseries 01}
  (2021) 018} [\href{http://arxiv.org/abs/2006.06461}{{\ttfamily
  arXiv:2006.06461}}].

\bibitem{BenAchour:2020fgy}
J.~Ben~Achour, H.~Liu, H.~Motohashi, S.~Mukohyama and K.~Noui, {\it {On
  rotating black holes in DHOST theories}},
  \href{https://doi.org/10.1088/1475-7516/2020/11/001}{JCAP {\bfseries 11}
  (2020) 001} [\href{http://arxiv.org/abs/2006.07245}{{\ttfamily
  arXiv:2006.07245}}].

\bibitem{LIGOScientific:2016aoc}
{\scshape LIGO Scientific, Virgo} collaboration, B.~P. Abbott et~al., {\it
  {Observation of Gravitational Waves from a Binary Black Hole Merger}},
  \href{https://doi.org/10.1103/PhysRevLett.116.061102}{Phys. Rev. Lett.
  {\bfseries 116} (2016) 061102}
  [\href{http://arxiv.org/abs/1602.03837}{{\ttfamily arXiv:1602.03837}}].

\bibitem{LIGOScientific:2016lio}
{\scshape LIGO Scientific, Virgo} collaboration, B.~P. Abbott et~al., {\it
  {Tests of general relativity with GW150914}},
  \href{https://doi.org/10.1103/PhysRevLett.116.221101}{Phys. Rev. Lett.
  {\bfseries 116} (2016) 221101}
  [\href{http://arxiv.org/abs/1602.03841}{{\ttfamily arXiv:1602.03841}}].

\bibitem{LIGOScientific:2020tif}
{\scshape LIGO Scientific, Virgo} collaboration, R.~Abbott et~al., {\it {Tests
  of general relativity with binary black holes from the second LIGO-Virgo
  gravitational-wave transient catalog}},
  \href{https://doi.org/10.1103/PhysRevD.103.122002}{Phys. Rev. D {\bfseries
  103} (2021) 122002} [\href{http://arxiv.org/abs/2010.14529}{{\ttfamily
  arXiv:2010.14529}}].

\bibitem{LIGOScientific:2021sio}
{\scshape LIGO Scientific, VIRGO, KAGRA} collaboration, R.~Abbott et~al., {\it
  {Tests of General Relativity with GWTC-3}},
  \href{https://doi.org/10.1103/PhysRevD.112.084080}{Phys. Rev. D {\bfseries
  112} (2025) 084080} [\href{http://arxiv.org/abs/2112.06861}{{\ttfamily
  arXiv:2112.06861}}].

\bibitem{EventHorizonTelescope:2019dse}
{\scshape Event Horizon Telescope} collaboration, K.~Akiyama et~al., {\it
  {First M87 Event Horizon Telescope Results. I. The Shadow of the Supermassive
  Black Hole}}, \href{https://doi.org/10.3847/2041-8213/ab0ec7}{Astrophys. J.
  Lett. {\bfseries 875} (2019) L1}
  [\href{http://arxiv.org/abs/1906.11238}{{\ttfamily arXiv:1906.11238}}].

\bibitem{EventHorizonTelescope:2022wkp}
{\scshape Event Horizon Telescope} collaboration, K.~Akiyama et~al., {\it
  {First Sagittarius A* Event Horizon Telescope Results. I. The Shadow of the
  Supermassive Black Hole in the Center of the Milky Way}},
  \href{https://doi.org/10.3847/2041-8213/ac6674}{Astrophys. J. Lett.
  {\bfseries 930} (2022) L12}
  [\href{http://arxiv.org/abs/2311.08680}{{\ttfamily arXiv:2311.08680}}].

\bibitem{Chandrasekhar1983}
S.~Chandrasekhar, \emph{The Mathematical Theory of Black Holes}. Oxford
  University Press, New York, 1983.

\bibitem{MisnerThorneWheeler1973}
C.~W. Misner, K.~S. Thorne and J.~A. Wheeler, \emph{Gravitation}. W. H.
  Freeman, San Francisco, 1973.

\bibitem{HackmannXu2013}
E.~Hackmann and H.~Xu, {\it Charged particle motion in kerr-newman
  space-times}, \href{https://doi.org/10.1103/PhysRevD.87.124030}{Physical
  Review D {\bfseries 87} (2013) 124030}
  [\href{http://arxiv.org/abs/1304.2142}{{\ttfamily arXiv:1304.2142}}].

\bibitem{Carter1968}
B.~Carter, {\it Global structure of the kerr family of gravitational fields},
  \href{https://doi.org/10.1103/PhysRev.174.1559}{Physical Review {\bfseries
  174} (1968) 1559}.

\end{thebibliography}\endgroup
	
\end{document}